# Investigation of the magnetic field effects on the electron mobility in tri-(8-hydroxyquinoline)-aluminum based light-emitting devices

Qiming Peng, Jixiang Sun, Xianjie Li, Mingliang Li, Feng Li[a)]

State Key Lab of Supramolecular Structure and Materials, Jilin University, 2699 Qianjin Avenue, Changchun 130012, People's Republic of China

**Abstract**: We investigated the mganetic field effects (MFEs) on electron mobility in tri-(8-hydroxyquinoline)-aluminum based light-emitting devices by transient-electroluminescence method upon application of various offset voltages. It is found the rising edges of EL pulses are well overlapped and the falling edges of EL pulses are separated for the magnetic field on and off when $V_{offset}=0\ V$ and $V_{offset} > V_{turnon}$ of the devices. The results suggest the bipolaron model and the triplet-polaron interaction model related to the carriers mobilities are not the dominant mechanisms to explain the MFEs under our experimental conditions, and the magnetic field affects the carriers recombination process is confirmed.

[a)] Author to whom correspondence should be addressed; Electronic mail: lifeng01@jlu.edu.cn

Since the first observation of magnetic field effects (MFEs) on emission and current in organic light-emitting diodes (OLEDs) with nonmagnetic electrodes in 2003 was reported by kalinowski *et al.*,[1] the MFEs on OLEDs have been studied extensively[2-27] due to the potential to enhance the efficiency of organic devices or produce the next-generation semiconductor device.[7] Despite the large progress in the research of MFEs on organic light-emitting diodes, the microscopic mechanism still deserves further investigation. To the current, a number of models have been proposed to explain the MFEs.[1,4,6,9,13,15,24,28] In these models, there are two mechanisms that are correlated to the carrier mobilities. One is the bipolaron model[15] suggesting that the mobilities are a function of inter-charge (or inter-polaron) spin interaction which can be affected by the magnetic field. Our previous results demonstrate that the bipolaron model is not the dominant mechanism in our devices under the pulse driving voltage.[17] The other is the triplet-polaron interaction (TPI) model[28] suggesting that the TPI could reduce the carriers mobilities and since the external magnetic field could shift the ratio of the triplets to singlets thus the magnetic field can affect the carrier mobilities. Recently, Song *et al.*[13] reported their results about the MFEs on the hole mobility (TPI model) in films of N,N'-diphenyl-N,N'-bis(3-methylphenyl)- (1,1'-biphenyl)-4,4'-diamine (TPD) by the dark injection method.[29] Motivated by their works, we carried out this experiment to investigate the MFEs on the electron mobility in tri-(8-hydroxyquinoline)-aluminum ($Alq_3$) based light-emitting diodes and test whether the TPI model is appropriate to interpret the MFEs in our experimental conditions.

In this work, a transient electroluminescence (EL) method was used.[30] Similar with the work of Song *et al.*,[13,29] an offset voltage ($V_{offset}$) was added to the pulse voltage ($V_{pulse}$) to test the TPI. The detailed definition of $V_{offset}$ , $V_{pulse}$ and $V_{total}$ is shown in inset of Fig. 4. Different with the dark injection method which acquires the transient current signals, the transient EL method in our experiment acquires the transient EL signals. We used an Agilent 8114A pulse generator (100v/2A) to apply rectangular pulses with repetition of 1 kHz and width of 7 μs or 10 μs to our devices, and the $V_{offset}$ were obtained by setting the baseline nonzero. The devices were fixed to a teflon stage with the magnetic field from an electrical magnet perpendicular to its current direction. The light emission was collected by a lens coupled with an optic fiber connected to a Hamamatsu photomultiplier (H5783-01) with time resolution of 0.78 ns. Then the photomultiplier was connected to a digital oscilloscope (Tektronix DPO7104, band width of 1 GHz) with the input resistance as 50Ω. The RC time of the circuit was less than 0.1μs. All measurements were carried out at room temperature under ambient condition.

The structure of the devices was ITO/ N, N′-di-1-naphthyl-N, N′-diphenylbenzidine (NPB) (60 nm)/tri-(8-hydroxyquinoline)-aluminum ($Alq_3$) (80 nm)/LiF (0.8 nm)/Al. The NPB was the hole-transport layer while the $Alq_3$ functioned as both emissive and electron-transport layer. Because the hole mobility in NPB layer is at least one order of magnitude larger than the electron mobility in $Alq_3$ layer. The mobility of electron in $Alq_3$ layer can be calculated by $\mu_e=L/(t_dE)$, approximately, where $L$ is the thickness of $Alq_3$ layer, $E$ is the electric field in $Alq_3$ layer and $t_d$ is the delay time from the moment when the pulse voltage is applied to the device to the appearance of the rising edge of the pulse EL signal.[30] Fig. 1 shows the transient EL response of the device to the magnetic field (150 mT) on and off driven by different $V_{pulse}$ with the $V_{offset}$, that is the baseline, is zero. As can be seen, the rising edges are well overlapped and the falling edges are separated from each other for the curves with the magnetic field on and off. In addition, the magnetoelectroluminescence (MEL) defined as $MEL=(EL(B) - EL(0))/EL(0)$ turns from positive to negative for the EL pulses at flat period when the pulse voltage increases from 7 V to 18 V. These results are consistent with our previous work[17,18] and suggest that the bipolaron model is not the dominant mechanism in our devices.

What we are more interested is the transient EL responses with the magnetic filed on and off under the condition that the $V_{offset}$ is larger that the turn-on voltage ($V_{turnon}$) of the device. When the $V_{offset}$ is larger than the $V_{turnon}$, the triplets should be pre-exist in $Alq_3$ layer before the pulse voltage arrives because of the ambipolar-current injection and the long lifetime of triplet excitons. Assuming the triplets could reduce the carrier mobility by TPI, the mobility of electron in $Alq_3$ layer with magnetic field on should larger than that with magnetic field off owing to that the magnetic field could decrease the concentration of triplets. So the obtained EL pulses with magnetic field on should ahead of those with magnetic field off, leading to the rising edges of EL pulse are separated with the magnetic field on and off. Thus whether the $t_d$ is changed with the magnetic on and off under the condition that the $V_{offset} > V_{turnon}$ can be used to check the TPI mode. Fig. 2 shows the transient EL responses under the condition that the $V_{offset} > V_{turnon}$ of the device. The EL signals can be observed before the pulse voltage is applied, meaning the ambipolar current-injection and the creation of the excitons. The values of MELs are positive and decline with the $V_{offset}$ increases from 5 V to 7 V. After $V_{pulse}$ of 17 V is applied, the MEL changes to be negative. The positive and negative MELs are observed in the time regions of $V_{offset}$ and $V_{pulse}$, respectively, in one EL pulse tested with the magnetic on and off. This further convinces our previous result that the values of MELs decline with the increase of driving voltage and turn from positive to negative sign when the driving voltage is big enough.[18]

Fig. 3 depicts the normalized transient EL responses of the device driven by $V_{pulse}$ from 3 V to 19 V and $V_{offset}$ of 5 V with the magnetic field on and off. As can be seen, the rising edges of the EL pulses with the magnetic field on and off overlap perfectly. This suggests that the $t_d$ is not changed by the magnetic field. Thus it can be inferred that the magnetic field does not affect the electron mobility in $Alq_3$ layer when the $Alq_3$ is full of triplet exctions. That is, the TPI mode is not the dominant mechanism for the MFEs under our experimental conditions. On the contrary, the falling edges of the EL pulses are separated with the magnetic field on and off, which confirms that the magnetic field actually affects the carriers recombination process.[9,17] The same conclusion can be drawn when the offset voltage changes from 4 V to 7 V.

Fig. 4 shows the normalized transient EL responses of the device upon application of various offset voltages from 1 V to 9 V, and the total voltage is kept constant as 14 V. If the TPI plays a major role, the electron drift velocity should drop because of the site blocking effect of triplet excitons when $V_{offset}$ is larger than $V_{turnon}$ (～2.5 V here), that is, the ambipolar injection is achieved and the triplet excitons are created. This would induce that the $t_d$ increase with the $V_{offset}$. However, from Fig.4 we can see that the $t_d$ decreases with the $V_{offset}$ rather than increases. This further verifies that the TPI mode is not the dominant mechanism for the MFEs under our experimental conditions. Here the $t_d$ reflects the time when the two leading fronts of injected carriers, electrons and holes, meet in $Alq_3$ layer.[30] The holes in $Alq_3$ layer are minority carrier and the penetrating distance of holes in $Alq_3$ layer increases with the $V_{offset}$, which means the meeting position of the leading fronts of electrons and holes moves toward the cathode with the increase of the $V_{offset}$, thus inducing the decrease of $t_d$.

We changed the light-emitting layer of the device from $Alq_3$ to $Alq_3$ doped Rubrene (2 wt %) and $Alq_3$ doped DCJTB (4-(Dicyanomethylene)-2-tert-butyl-6-(1,1,7,7-tetramethyljulolidin-4-yl-vinyl)-4H-pyran) (2 wt %). The similar results can be obtained.

In summary, we investigated the MFEs on electron mobility in $Alq_3$ based light-emitting devices by transient EL method upon application of various offset voltages. It is found that the rising edges of EL pulses are well overlapped and the falling edges of EL pulses are separated for the magnetic field on and off when $V_{offset}=0\ V$ and $V_{offset} > V_{turnon}$ of the devices. The results implies that the magnetic field has no effect on electron mobility in $Alq_3$ under the condition of $V_{offset}=0\ V$ and $V_{offset} > V_{turnon}$. Thus the bipolaron model and the TPI model related to the carriers mobilities are not the dominant mechanisms to explain the MFE under our experimental conditions, and the magnetic field affects the carriers recombination process is confirmed.

We are grateful for financial support from National Natural Science Foundation of China (grant numbers 60878013 and 60706016).

**Figure captions**

**FIG.1.** The transient EL responses of the device driven by various rectangular pulse voltage (the repetition and the width are 1 KHz and 7 μs, respectively. The offset voltage is zero) with the magnetic field of 150 mT on (red line) and off (black line). (a), (b), (c) and (d) shows the details of the EL curves at rising edges, flat period and falling edges, respectively.

**FIG.2.** The transient EL responses of the device driven by rectangular pulse voltage of 17 V (the repetition and the width are 1 KHz and 10 μs, respectively, and the offset voltage increases from 5 V to 7 V) with the magnetic field of 150 mT on (red line) and off (black line). The values of MELs correlated to the offset voltages and the pulse voltages are shown in the texts, respectively.

**FIG.3.** The normalized transient EL responses of the device driven by rectangular pulse voltage from 3 V to 19 V (the repetition and the width are 1 KHz and 10 μs, respectively, and the offset voltage is 5 V) with the magnetic field of 150 mT on (red line) and off (black line).

**FIG.4.** The normalized transient EL responses of the device upon application of various offset voltages from 1 V to 9 V. The total voltage is kept constant as 14 V. Insert: the sketch of $V_{offset}$, $V_{pulse}$ and $V_{total}$, respectively.

FIG.1.

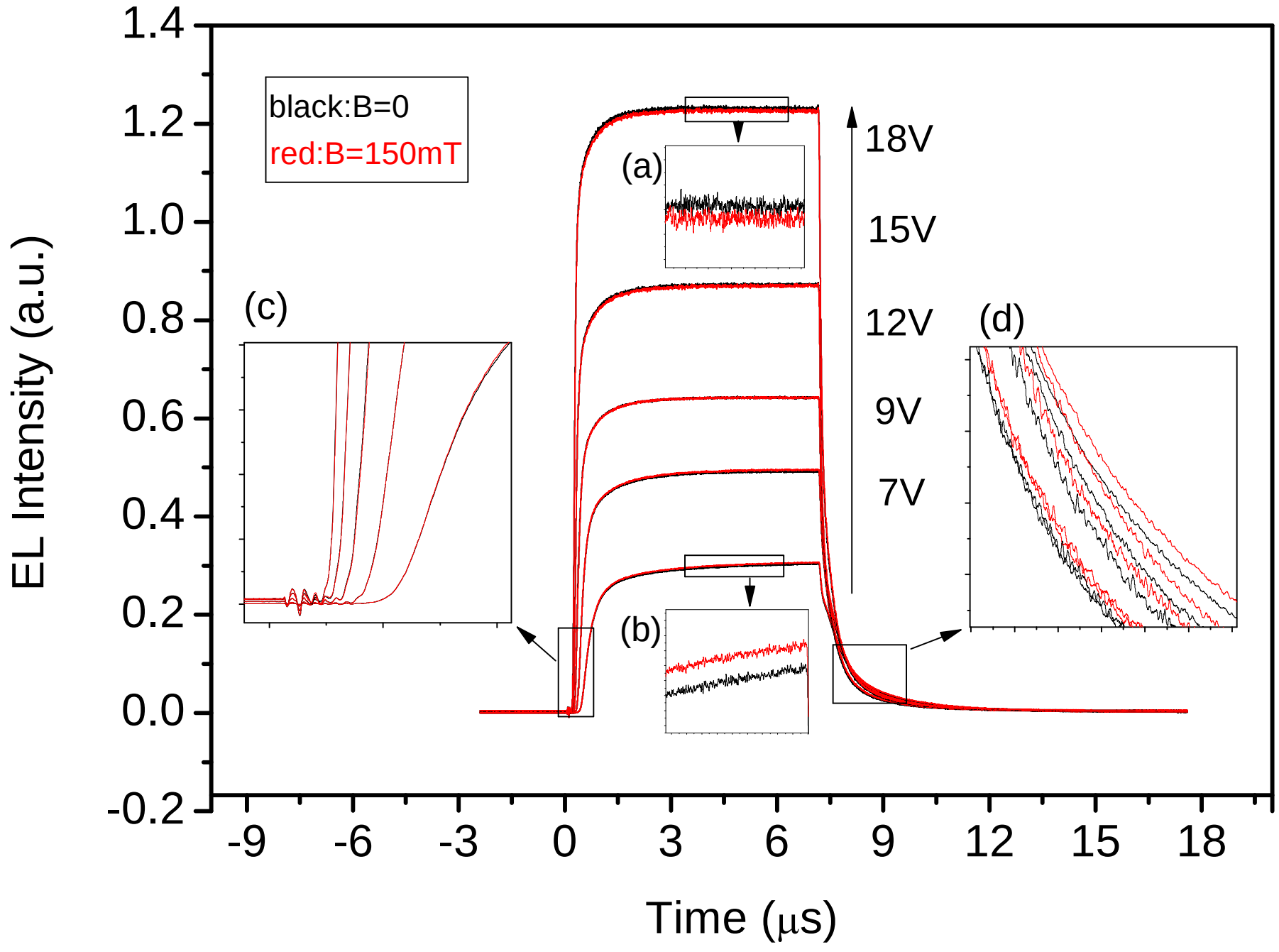

FIG.2.

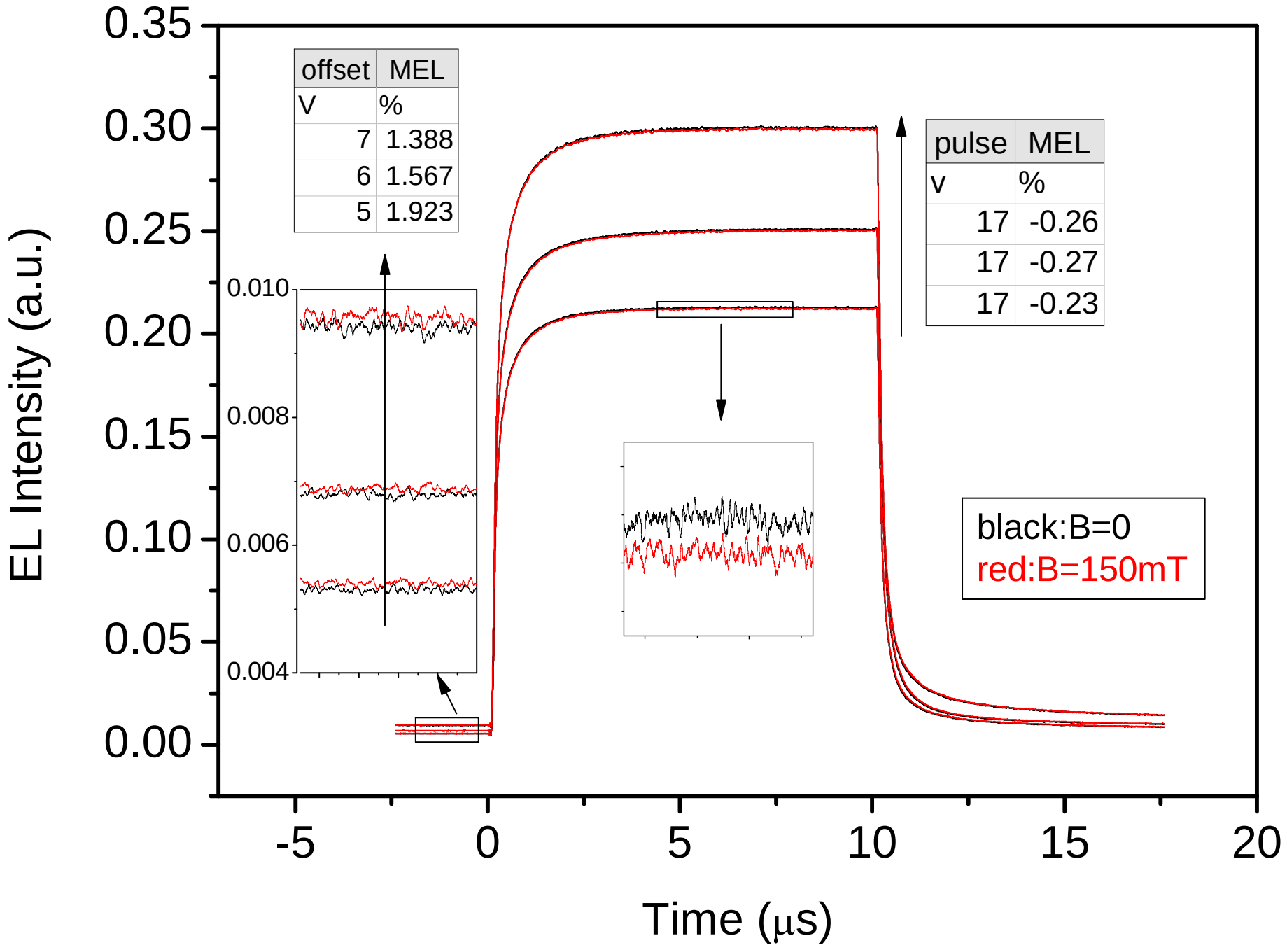

FIG.3.

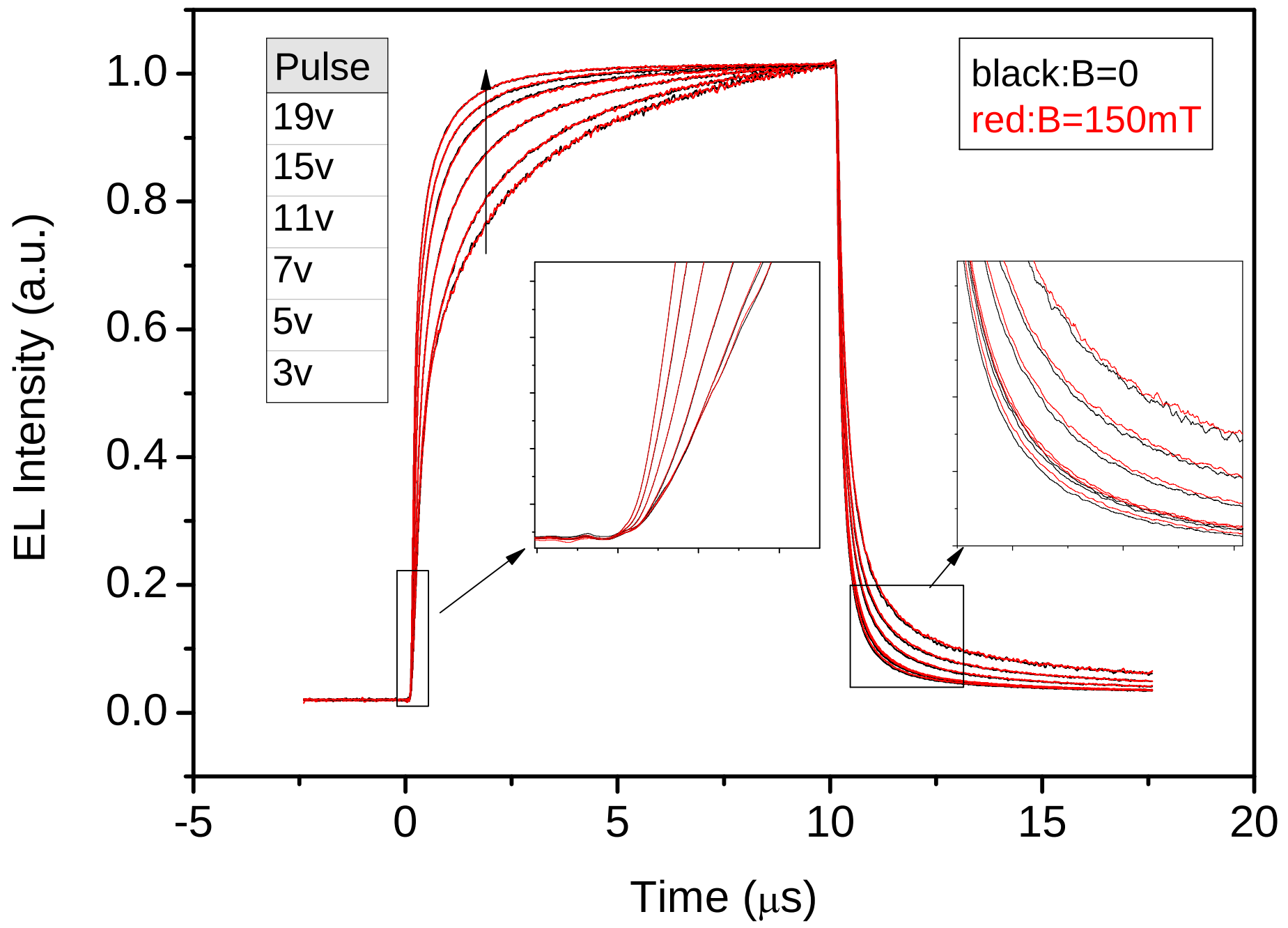

FIG.4.

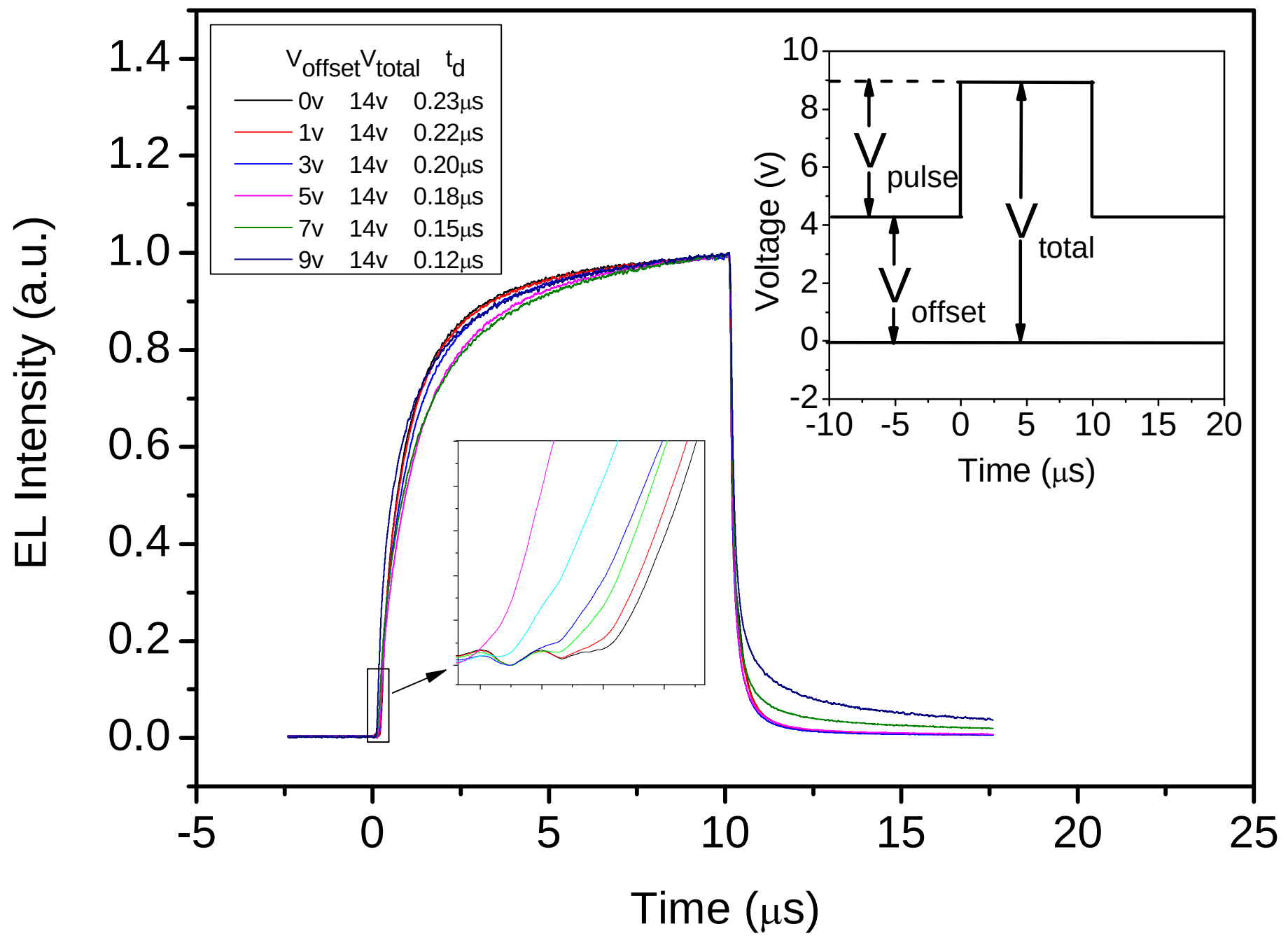